\documentclass[aps,pra,showpacs,preprint]{revtex4-1}
\usepackage{amsmath}
\usepackage{graphicx}% Include figure files %[dvipdfmx] for revtex
\usepackage{dcolumn}% Align table columns on decimal point
\usepackage{bm}% bold math
\usepackage{textcomp}
\usepackage{natbib}
\usepackage{upgreek}
\usepackage{enumerate}
\usepackage{siunitx}

\def\bra#1{\mathinner{\langle{#1}|}}
\def\ket#1{\mathinner{|{#1}\rangle}}

\renewcommand\Re{\operatorname{Re}}

\begin{document}
\title{Study of EIT resonances in an anti-relaxation coated Rb vapor cell}
	
%\author[label1]{Mangesh Bhattarai}
%\author[label1]{Vineet Bharti}
%\author[label1]{Vasant Natarajan\corref{cor1}}
%%\ead{vasant@physics.iisc.ernet.in}
%%\ead[url]{http://www.physics.iisc.ernet.in/~vasant}
%\author[label2]{Armen Sargsyan}
%\author[label2]{David Sarkisyan}
%%\cortext[cor1]{Corresponding author}
%\address[label1]{Department of Physics, Indian Institute of Science, Bangalore-560012, India}
%\address[label2]{Institute for Physical Research, NAS of Armenia, Ashtarak-2, 0203, Armenia}
\author{Mangesh Bhattarai}
\author{Vineet Bharti}
\author{Vasant Natarajan}
\email{vasant@physics.iisc.ernet.in}
\affiliation{Department of Physics, Indian Institute of Science, Bangalore-560012, India}
\author{Armen Sargsyan}
\author{David Sarkisyan}
\affiliation{Institute for Physical Research, NAS of Armenia, Ashtarak-2, 0203, Armenia}

\begin{abstract}
	We demonstrate---experimentally and theoretically---that resonances obtained in electromagnetically induced transparency (EIT) can be both bright and dark. The experiments are done using magnetic sublevels of a hyperfine transition in the D$_1$ line of $^{87}$Rb. The degeneracy of the sublevels is removed by having a magnetic field of value 27 G. The atoms are contained in a room-temperature vapor cell with anti-relaxation coating on the walls. Theoretical analysis based on a two-region model reproduces the experimental spectrum quite well. This ability to have both bright and dark resonances promises applications in sub- and super-luminal propagation of light, and sensitive magnetometry.	
\end{abstract}

	
\maketitle	

\thispagestyle{empty}

\section{Introduction}

Phenomena involving optical coherences---coherent population trapping (CPT) \cite{ARI96}, electromagnetically induced transparency (EIT) \cite{HAR97,FIM05}, electromagnetically induced absorption (EIA) \cite{GWR04,KTW07}, Hanle effect \cite{KIM12,RBB17}, and so on---have been studied extensively by the quantum optics community. The main system used in these studies are alkali atoms like Rb and Cs. This is because these atoms can be housed in glass vapor cells, have a high vapor pressure at room temperature, and have transitions that are accessible with diode lasers. Coherences involving ground-state energy levels gain by the use of a buffer gas in the vapor cell, or anti-relaxation coating (ARC) like paraffin on the cell walls, because this increases the coherence time and results in narrowing of the resonances. Such narrow resonances have important applications in fields such as atomic clocks \cite{WYN99}, sensitive magnetometry \cite{BGK02}, transporting and storage of light for use in quantum memories \cite{ZMK02}, and the search for a permanent electric dipole moment in atoms \cite{RBA16}.

Many of these studies have been done using Zeeman sublevels of a particular hyperfine transition of the ground state in these atoms. This is because the frequency difference in the presence of a small magnetic field is small enough that it can be achieved with acousto-optic modulation, if phase coherence between the control and probe beams is required. In the absence of a magnetic field, the sublevels form one or more degenerate $\Lambda$-type systems. In the presence of a magnetic field, the degeneracy is lifted, which results in a narrowing of EIT resonances \cite{IFN09} and the creation of new CPT dark states \cite{MRW13a}. When the field is sufficiently strong, it has been shown that a dark resonance (DR) in EIT gets transformed to a bright resonance (BR) \cite{SSP15,SSM16}. The use of an ARC on the cell walls in these experiments improves the strength of the BR \cite{SSM16}.

In this work, we show that EIT resonances of both kinds---BRs and DRs---are obtained in a room-temperature Rb vapor cell in the presence of a magnetic field. The cell has ARC on the walls.  The experiments are done using magnetic sublevels of the ground state in the D$_1$ line ($ 5S_{1/2} \rightarrow 5P_{1/2}$ transition) of $^{87}$Rb at 795 nm. The experimental results are supported by a two-region theoretical model \cite{ROC10,BBN17} incorporating all the magnetic sublevels of the ground state.

This ability to transform between BRs to DRs shows that the medium can be switched from sub- to super-luminal propagation of light. In order to highlight this point, we show the variation of group refractive index with probe detuning. The separation between the resonances is proportional to the magnetic field, hence the technique can be used for sensitive magnetometry.

\section{Experimental details}

\begin{figure}
	\centering
	\includegraphics[width=.9\textwidth]{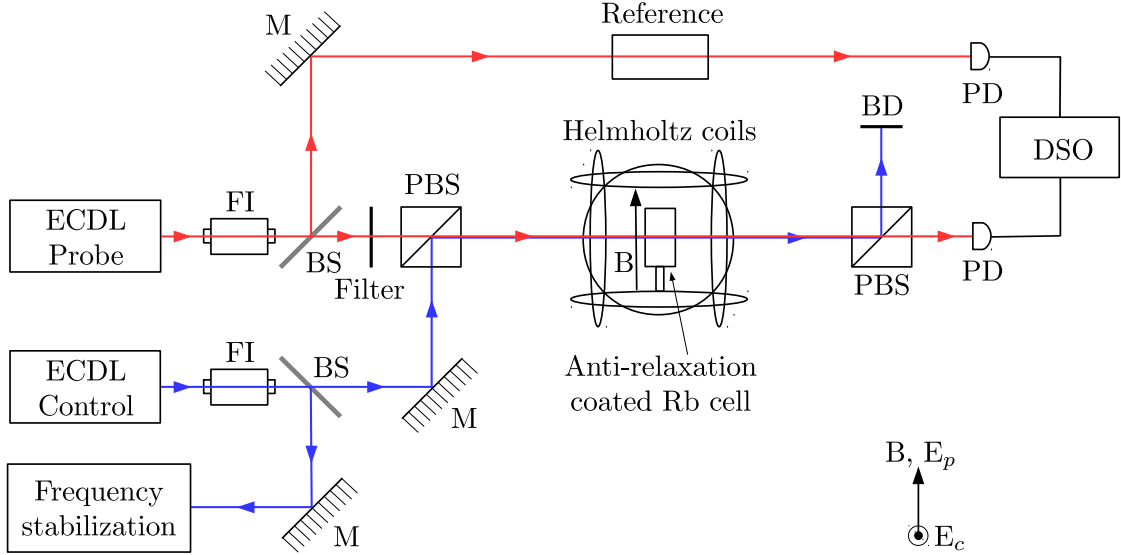}
	\caption{Experimental schematic. The direction of the B field with respect to the direction of polarizations of the probe and control beams is shown. Figure key: ECDL -- external cavity diode laser;  BS -- beam splitter; M -- mirror; FI -- Faraday isolator; PBS -- polarizing beam splitter; BD -- beam dump; PD -- photodiode; DSO -- digital storage oscilloscope.}
	\label{expt_setup}
\end{figure}

The experimental arrangement is shown schematically in Fig.~\ref{expt_setup}. The probe and control beams are derived from grating stabilized external cavity diode lasers (ECDLs) operating near the 795 nm D$_1$ line of Rb. The linewidth after stabilization is about 1 MHz. The beams (derived from independent laser systems) have orthogonal linear polarizations, and are mixed on a polarizing beam splitter (PBS) cube. The combined beams enter a vapor cell containing Rb and with ARC on the walls. The cell is inside three orthogonal pairs of Helmholtz coils. The  Helmholtz coils are used to nullify any stray magnetic field in the laboratory, and also apply a magnetic field by increasing the current in the corresponding pair. 

The power in the control beam is varied in the range of 1 to 30 mW, while that in the probe beam is varied from \SI{5}{\micro\watt} to 0.5 mW. The size ($1/e^2$ diameter) of the beams is about 1 mm. The control frequency is stabilized with a modified dichroic atomic vapor laser lock (DAVLL), as described in Refs.~\cite{SSP15} and \cite{SLP11}. The frequency reference spectrum is formed using the usual saturated absorption spectroscopy technique. The probe and control beams are separated using a second PBS, and only the probe beam is measured with a photodiode. The signal is recorded in a digital storage oscilloscope (DSO).

The vapor cell is at room temperature and has Rb in its natural abundance (72\% of $^{85}$Rb and 28\% of $^{87}$Rb). The inside surfaces (including windows) have ARC provided by a thin film of Polydimethylsiloxane (PDMS). PDMS was chosen because it is resistant to heating up to \SI{120}{\degreeCelsius}. The cell has dimensions of 25 mm diameter $\times$ 10 mm length.

\section{Theoretical analysis}
Theoretical analysis of the experimental spectra is done using a two-region model. The model is based on the multi-region density-matrix analysis given in Refs.~\cite{ROC10}  and \cite{BBN17}. The two interaction regions are labeled bright and dark. In the bright region both light and magnetic fields are present, while in the dark region there is no light and only a magnetic field.

\begin{figure}
	\centering
	\includegraphics[width=.8\textwidth]{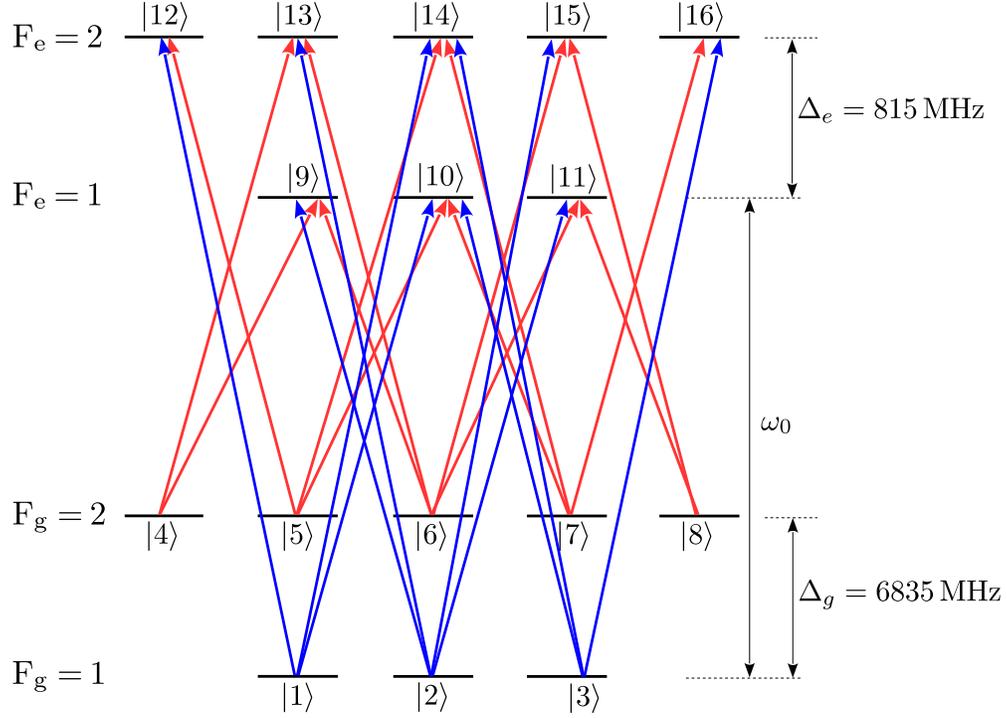}
	\caption{Magnetic sublevels of the relevant transition in $^{87}$Rb (not to scale). Transitions coupled by $\vec{E}_1$ field are shown in red, while transitions coupled by $\vec{E}_2$ field are shown in blue.}
	\label{levelsfortheory}
\end{figure}

We consider two linear and mutually perpendicular polarized light fields $\vec{E}_1$ and $\vec{E}_2$ with frequencies $\omega_1$ and $\omega_2$ traveling along $ \hat{z} $, expressed as
\begin{equation}
\begin{aligned}
\vec{E}_1 = E_{10}\cos(\omega_1 t) \,\hat{y}  \\ 
\vec{E}_2 = E_{20} \cos(\omega_1 t) \,\hat{x}
\end{aligned}
\end{equation}
where $E_{10}$ and $E_{20}$ are the respective amplitudes of the two fields. Both the fields are interacting with Rb in a vapor cell with ARC on walls. The field $\vec{E}_1$ is applied on the $F_g=1 \to F_e=1$ transition and the field $\vec{E}_2$ is applied on the $F_g=2 \to F_e=1$ transition of the D$_1$ line of $^{87}$Rb. The magnetic sublevels of this line and the transitions driven by the two fields are shown in Fig.~\ref{levelsfortheory}.

We consider that the ground state hyperfine splitting is $\Delta_g$, excited state hyperfine splitting is $\Delta_e$ and transition frequency of $F_g=1 \rightarrow F_e=1$ is $\omega_0$. The detunings of the light fields from the respective transitions are
\begin{equation}
\begin{aligned}
\delta_1 &=\omega_1 - \omega_0,\\ 
\delta_2 &=\omega_2 - (\omega_0-\Delta_g)
\end{aligned}
\end{equation}
In the rotating frame, the Hamiltonian describing the atomic states of the system is
\begin{equation}
\begin{aligned}
H_{0}&=(\delta_1-\delta_2)(\ket{4}\bra{4} +  \ket{5}\bra{5} +  \ket{6}\bra{6} +  \ket{7}\bra{7} +  \ket{8}\bra{8}) \\ 
&-\delta_2( \ket{9}\bra{9} +  \ket{10}\bra{10} +  \ket{11}\bra{11}) \\ 
&+ (\Delta_e-\delta_2) (\ket{12}\bra{12} +  \ket{13}\bra{13} +  \ket{14}\bra{14} + \ket{15}\bra{15} + \ket{16}\bra{16}) 
\label{Hatom} 
\end{aligned}
\end{equation} 
and the interaction Hamiltonian describing the light-atom interaction is
\begin{equation}
\begin{aligned}
H_{\rm int}&=\frac{\hbar \Omega_1}{2} \bigg[\ C_{1,10} \ket{1}\bra{10}+C_{1,12}\ket{1}\bra{12}+C_{1,14}\ket{1}\bra{14}\\
&\hspace{1.25cm}+C_{2,9}\ket{2}\bra{9}+C_{2,11}\ket{2}\bra{11}+C_{2,13}\ket{2}\bra{13}+C_{2,15}\ket{2}\bra{15}\\
&\hspace{1.25cm}+C_{3,10}\ket{3}\bra{10}+C_{3,14}\ket{3}\bra{14}+C_{3,16}\ket{3}\bra{16}\ \bigg] \\
&+\frac{\hbar \Omega_2}{2} \bigg[\ C_{4,9} \ket{4}\bra{9} + C_{4,13} \ket{4}\bra{13} + C_{5,10} \ket{5}\bra{10} + C_{5,12} \ket{5}\bra{12}  \\ 
&\hspace{1.25cm}+C_{5,14} \ket{5}\bra{14}+C_{6,9} \ket{6}\bra{9} + C_{6,11} \ket{6}\bra{11} + C_{6,13} \ket{6}\bra{13}\\ 
&\hspace{1.25cm}+ C_{6,15} \ket{6}\bra{15} +C_{7,10} \ket{7}\bra{10}  + C_{7,14} \ket{7}\bra{14} + C_{7,16} \ket{7}\bra{16} \\
&\hspace{1.25cm}+ C_{8,11} \ket{8}\bra{11} + C_{8,15} \ket{8}\bra{15}\ \bigg]   \\
&+{\rm Hermitian \, conjugate}  
\label{H_int}
\end{aligned}
\end{equation}
In Eq.~\eqref{H_int}, $C_{i,j}$s are the Clebsch-Gordan coefficients (along with a factor determined by the polarization of the beam) connecting the state $\ket{i}$ and $\ket{j}$, and are written in terms of $\bra{J_e=\frac{1}{2}}d\ket{J_g=\frac{1}{2}}$ as
\begin{equation*}
\begin{aligned}
&C_{1,10}=C_{1,14}=-C_{2,9}=C_{2,11}=-C_{3,10}=C_{3,14}=\dfrac{i}{2\sqrt{6}}\\
&C_{1,12}=C_{3,16}=\dfrac{i}{2}\\
&C_{2,13}=C_{2,15}=\dfrac{i}{2\sqrt{2}}\\
&C_{1,9} =-C_{3,11}=-C_{4,13}=-C_{5,12}=-C_{7,16}=-C_{8,15}=\dfrac{1}{2 \sqrt{3}}\\
&C_{1,13}=C_{3,15}=-C_{4,9}=C_{8,11}=\dfrac{1}{2}\\
&C_{2,14}=\dfrac{1}{\sqrt{3}}\\
&C_{5,10}=C_{5,14}=C_{6,13}=C_{6,15}=-C_{7,10}=C_{7,14}=-\dfrac{1}{2\sqrt{2}}\\
&C_{6,9}=-C_{6,11}=\dfrac{1}{2 \sqrt{6}}
\end{aligned}
\end{equation*}
In addition to the light fields, a magnetic field is applied to the system along the direction of polarization of the probe beam. The magnetic field Hamiltonian is 
\begin{equation}
\begin{aligned}
H_{B} &= \dfrac{\mu_B B_x}{\hbar} \left[g_{F_g} \sum_{i,j = 1}^{3} \sigma^{x(3)}_{i,j} + g_{F_{g'}} \sum_{i,j = 4}^{8} \sigma^{x(5)}_{i,j} + g_{F_e} \sum_{i,j = 9}^{11} \sigma^{x(3)}_{i,j} + g_{F_{e'}} \sum_{i,j = 12}^{16} \sigma^{x(5)}_{i,j}\right] \\ 
&+\dfrac{\mu_B B_y}{\hbar} \left[g_{F_g} \sum_{i,j = 1}^{3} \sigma^{y(3)}_{i,j} + g_{F_{g'}} \sum_{i,j = 4}^{8} \sigma^{y(5)}_{i,j} + g_{F_e} \sum_{i,j = 9}^{11} \sigma^{y(3)}_{i,j} + g_{F_{e'}} \sum_{i,j = 12}^{16} \sigma^{y(5)}_{i,j}\right]
\label{HB}
\end{aligned}
\end{equation}
In Eq.~\eqref{HB}: $B_x$ and $B_y$ are the the magnetic field along $x$ and $y$ directions; $g_{F_g}$ is the Lande $g$ factor of the ground $F=1$ level, $g_{F_{g^{\prime}}}$ is the Lande $g$ factor of the ground $F=2$ level, $g_{F_e}$ is the Lande $g$ factor of the excited $F=1$ level, and $g_{F_{e^{\prime}}}$ is the Lande $g$ factor of the excited $F=2$ level; $\mu_B$ is Bohr magneton; and $\sigma^{x(k)}$ and $\sigma^{y(k)}$ (with $k=2F+1$) are the Pauli matrices for spin $F$ particles. 

The Hamiltonians in the bright ($H_{\rm b}$) and dark ($H_{\rm d}$) regions are written in terms of Hamiltonians for the free evolution ($H_0$), light-atom interaction ($H_{\rm int}$), and magnetic field interaction ($H_B$). Hence they are
\begin{equation}
\begin{aligned}
H_{\rm b} &= H_0 + H_{\rm int} + H_{\rm B} \\
H_{\rm d} &= H_0 + H_{\rm B}
\end{aligned}
\end{equation}
The time evolution of the system in two regions is given by following density matrix equations
\begin{equation}
\begin{aligned}
{\dot\rho^{\rm b}} &= -\frac{i}{\hbar} [H_{\rm b},\rho^{\rm b}] - \frac{1}{2} \{\rho^{\rm b},\hat{\Gamma}\} +\Lambda_{\rm b} - \gamma_{\rm b} \rho^{\rm b} + \gamma_{\rm b} \rho^{\rm d} \\
{\dot\rho^{\rm d}} &= -\frac{i}{\hbar} [H_{\rm d},\rho^{\rm d}] - \frac{1}{2} \{\rho^{\rm d},\hat{\Gamma}\} +\Lambda_{\rm d} - (\gamma_{\rm d} + \gamma_{\rm w})\rho^{\rm d} + \gamma_{\rm d} \rho^{\rm b} + \gamma_{\rm w} \Lambda^{\rm  u} 
\label{rhoequations}
\end{aligned}
\end{equation}
where $\rho^{\rm b}$ and $\rho^{\rm d}$ are the density matrices in the bright and dark regions, $\hat{\Gamma}$ is the relaxation matrix, $\Lambda_{\rm b}$ and $\Lambda_{\rm d}$ are the repopulation matrices for sublevels in the two ground levels, $\gamma_{\rm b}$ and $\gamma_{\rm d}$ are the transit-time decay rates for traversing the beams in the two regions, $\gamma_{\rm w}$ is the relaxation rate due to wall collisions, and $\Lambda^{\rm u}$ is the repopulation matrix for the sublevels in the two ground levels due to inter-atomic collisions.

$\hat{\Gamma}$ is a diagonal matrix with 0 for the two ground levels, and $\Gamma_e$ for the two levels in the excited state. $\Lambda_{\rm b}$ and $\Lambda_{\rm b}$ are diagonal matrices for repopulation due to spontaneous emission from the excited state along with branching ratios for each level. $\gamma_{\rm b}$ is taken to be the inverse of the time it takes for an atom with average velocity (at the cell temperature) to traverse the beam in the bright region. $\gamma_{\rm d}$ is taken to be proportional to $\gamma_{\rm b}$, with the proportionality factor given by the ratio of the volumes in the dark and bright regions. $\Lambda^{\rm u}$ is an unpolarized repopulation term, which for particular sublevels $\ket{i}$ and $\ket{j}$ belonging to a level $F$ is given by 
\begin{equation*}
\Lambda^{\rm u}_{i,i} = \dfrac{1}{2 F+1} \sum_{j} \rho^{\rm d}_{j,j}
\end{equation*}

In a cell with ARC, atoms in the dark region undergo numerous wall collisions. This process preserves the atomic polarization but randomizes the atomic velocity. Therefore, an atom that leaves the bright region re-enters with random velocity. This is called complete mixing of velocities \cite{ABR10}. To incorporate this process in our model, we solve the density matrix equation in two steps. First, we solve the density matrix equations given in Eq.~\eqref{rhoequations} to obtain the dark region density matrix as a function of velocity $v$ and detuning $\delta$. The complete velocity mixing assumption is then incorporated by performing an average over Maxwell-Boltzmann equation of the $\rho^{\rm d}$ as 
\begin{equation}
\begin{aligned}
\tilde{\rho}^{\rm d}_{ij} (\delta)&=\sqrt{\dfrac{m}{2 k_B T}} \int {\rm d}v  ~\rho^{\rm d}_{ij}(\delta,v)  ~\exp\left(-\dfrac{m v^2}{2 k_B T}\right)
\label{MBD}
\end{aligned}
\end{equation}
We now use the density matrix obtained from Eq.~\eqref{MBD} as the transit repopulation term in the density matrix for the bright region. With this modification, the equation of motion for the density matrix in the bright region is given as
\begin{equation}
\begin{aligned}
{\dot\rho^{\rm b}} &= -\frac{i}{\hbar} [H_{\rm b},\rho^{\rm b}] - \frac{1}{2} \{\rho^{\rm b},\hat{\Gamma}\} +\Lambda_{\rm b} - \gamma_{\rm b} \rho^{\rm b} + \gamma_{\rm b} \tilde{\rho}^{\rm d}\\
\label{rhobright_modified}
\end{aligned}
\end{equation}

\section{Results and discussion}

\subsection{First $\Lambda$-type system} 

\noindent
\textbf{The control laser is on the $F_g=1 \rightarrow F_e=1$ transition and the probe laser is scanned across the  $F_g=2 \rightarrow F_e=1$ transition.}

\begin{figure}
	\centering
	\includegraphics[width=.5\textwidth]{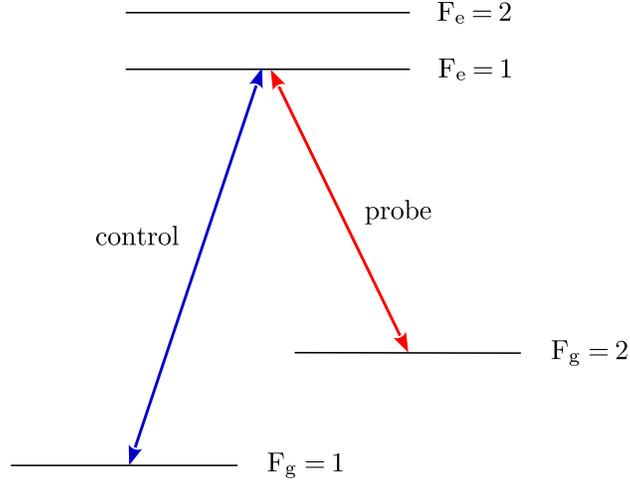}
	\caption{Levels coupled by the probe and control beams showing how the $\Lambda$-type system is formed.}
	\label{firstlambda}
\end{figure}

The levels coupled by the probe and control beams to form the first $\Lambda$-type system are shown in Fig.~\ref{firstlambda}. An external magnetic field whose direction (as indicated in the experimental schematic) is parallel to the polarization of the probe beam and orthogonal to that of the control beam.

\subsubsection{Experimental results}

\begin{figure}
	\centering
	\includegraphics[width=.6\textwidth]{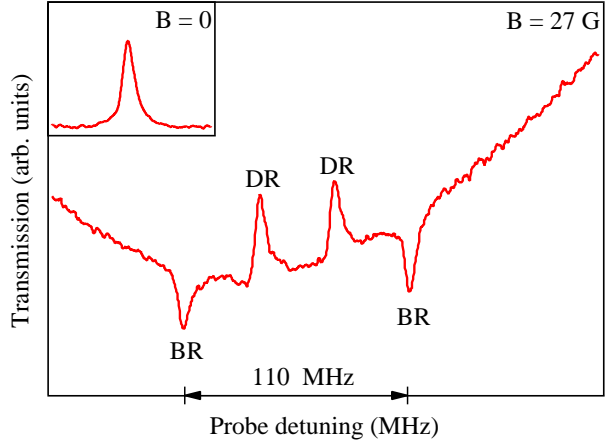} 
	\caption{Experimental spectrum of probe transmission showing both BRs and DRs. The curve is obtained with a probe power of \SI{10}{\micro\watt}, control power of 13.6 mW, and a magnetic field of 27 G. The inset shows usual EIT resonance in the absence of a field.}
	\label{result_firstlambda}
\end{figure}

The experimental spectrum is shown in Fig.~\ref{result_firstlambda}. The curve is obtained with a probe power of \SI{10}{\micro\watt} and a control power of 13.6 mW. The experiments are done in the presence of a magnetic field of 27 G and the cell is maintained at a temperature of \SI{36}{\degreeCelsius}.  The curve shows 2 DRs on the inside (corresponding to reduced EIT absorption), and 2 BRs on the outside (corresponding to enhanced EIT absorption). The inset shows that the usual EIT resonance is obtained in the absence of a field.

\subsubsection{Theoretical results}

\begin{figure}
	\centering
	\includegraphics[width=.6\textwidth]{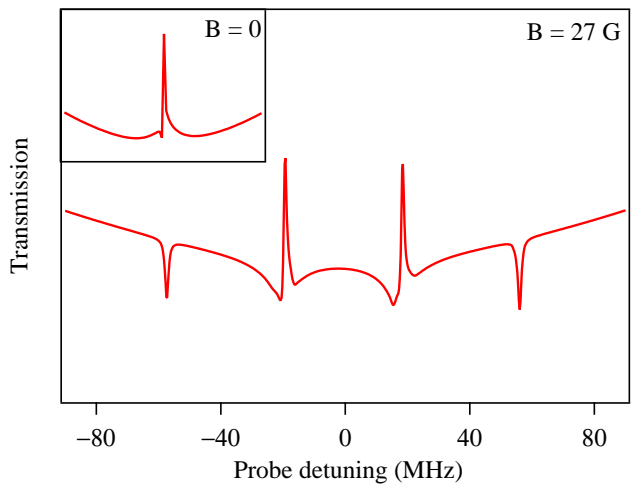}
	\caption{Calculated spectrum for the first $\Lambda$-type system. The Rabi frequency for the probe beam is $ 0.38 \, \Gamma_e$ (corresponding to the experimental value of \SI{10}{\micro \watt}), and that of the control beam is $13.9 \, \Gamma_e$ (corresponding to the experimental value of 13.6 mW). The values of the relaxation rates used are $\gamma_{\rm b} = 0.035 \,\Gamma_e$, $\gamma_{\rm d} = 0.000056 \,\Gamma_e$ , $\gamma_{\rm w} = 0.000174 \,\Gamma_e$. The inset shows that a single peak is obtained for 0 field.}
	\label{firstlambda_theory}
\end{figure}

The calculated spectrum is shown in Fig.~\ref{firstlambda_theory}. In this case, the probe beam corresponds to the $\vec{E}_2$ field, while the control beam corresponds to the $\vec{E}_1$ field. The calculation is done with probe and control Rabi frequencies corresponding to the powers used in the experiment. The power $P$ is related to the Rabi frequency $\Omega_R$ as 
\begin{equation*}
\Omega_R = \Gamma_e \sqrt{\dfrac{P}{\pi w_{0}^2} \dfrac{1}{2I_s}}
\end{equation*}
where $\Gamma_e = 2 \pi \times 5.89$ MHz is the natural linewidth of the excited state, $w_{0}$ is the radius of the beam, and $I_s = 4.484$ mW/cm$^2$ is the saturation intensity (the intensity at which the transition gets power broadened by a factor of $ \sqrt{2}$)  of the excited state. The intensity at the center of a Gaussian beam is actually twice as high, but this is taken as some kind of uniform value over the size of the beam. The values of the different relaxation rates are given in the figure caption. The calculation is done with Doppler averaging corresponding to the Maxwell-Boltzmann distribution at a temperature of \SI{36}{\degreeCelsius} (absolute temperature of 309 K). The calculation also reproduces the observed fact that a single peak is obtained when the magnetic field is 0, as seen from the inset of the figure.

What is not reproduced by theory in comparison to experiment are the following.
\begin{enumerate}
	\item The linewidth of the different resonances. This is probably because the Rabi frequency of the two beams are taken to be constant, while in reality they fall off like a Gaussian.
	
	\item Asymmetry in the absorption spectrum. This is probably due to the presence of nearby hyperfine levels in the excited state \cite{MSL11,BHW13}.
\end{enumerate}

A qualitative way of understanding the above complete theoretical analysis is that the enhancement or suppression of EIT resonances occurs because the population in the ground sublevels is different from that in thermal equilibrium, mainly due to optical pumping by the strong control beam \cite{SSP15,SSM16}.

\subsubsection{Control over group velocity}

This result is important for sub- and super-luminal velocity of light in the medium. In order to see this explicitly, we evaluate the group refractive index $n_g$ for group velocity $v_g$ in terms of the susceptibility $\chi$. It is given by \cite{BHN17}
\begin{equation*}
v_{g}=\frac{c}{n_g}=\frac{c}{1+\dfrac{\Re\{\chi\}}{2}+\dfrac{\omega_{p}}{2}\dfrac{\partial \Re\{\chi\}}{\partial\omega_{p}}} \quad , \quad \chi=\frac{N|\mu_{p}|^2}{\hbar\varepsilon_{0}\Omega_{p}}\rho_{eg}
\label{groupvelocity}
\end{equation*}
where $c$ is the velocity of light in vacuum, $\omega_p$ is the probe frequency, $N$ is the number density of atoms, $\mu_p$ is the dipole matrix element of the probe transition, and $\Omega_p$ is the probe Rabi frequency.

\begin{figure}
	\centering
	\includegraphics[width=.6\textwidth]{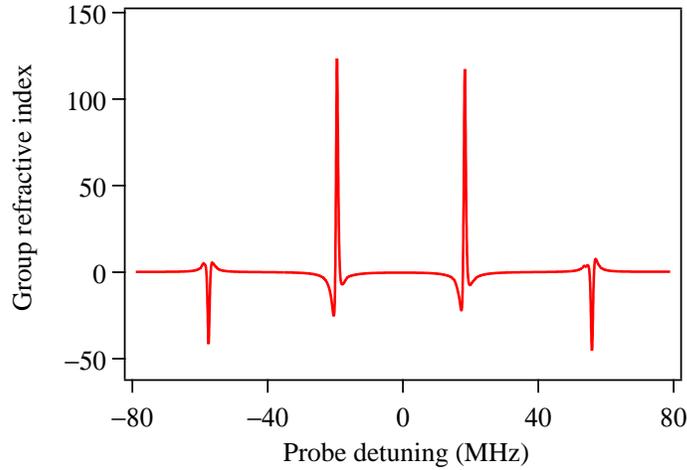}
	\caption{Calculated value of the group refractive index for the spectrum shown in Fig.~\ref{firstlambda_theory}.}
	\label{subsuper}
\end{figure}

We now plot the group refractive index as a function of probe detuning for the first $\Lambda$-type system. The results are shown in Fig.~\ref{subsuper}. As seen, the index is positive for DRs, which results in sub-luminal propagation of light \cite{HHD99}. BRs have negative index, which results in super-luminal propagation of light \cite{WKD00}. Thus, by tuning from bright to dark resonances, one can switch the system from sub-luminal to super-luminal propagation of light.

\subsection{Second $\Lambda$-type system}

\noindent
\textbf{The control laser is on the $F_g=2 \rightarrow F_e=1$ transition and the probe laser is scanned across the  $F_g=1 \rightarrow F_e=1$ transition.}

\begin{figure}
	\centering
	\includegraphics[width=.5\textwidth]{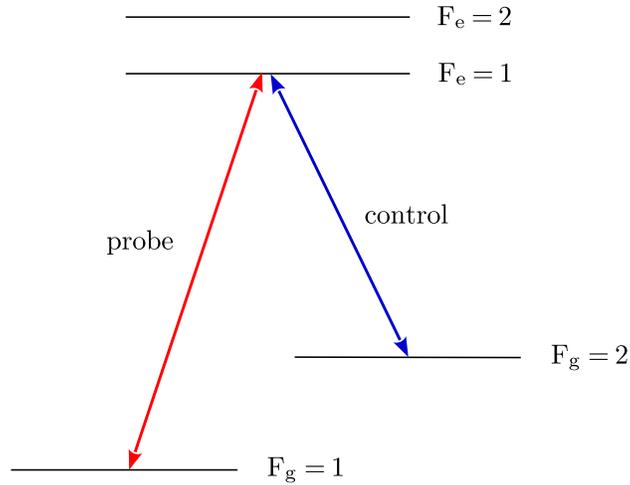}
	\caption{Levels coupled by the probe and control beams showing how the $\Lambda$-type system is formed.}
	\label{secondlambda}
\end{figure}

For the second $\Lambda$-type system, the roles of the probe and control beams are interchanged. The levels coupled by the two beams for this system are shown in Fig.~\ref{secondlambda}. A magnetic field with direction similar to the one before is applied.

\subsubsection{Experimental results}

\begin{figure}
	\centering
	\includegraphics[width=.6\textwidth]{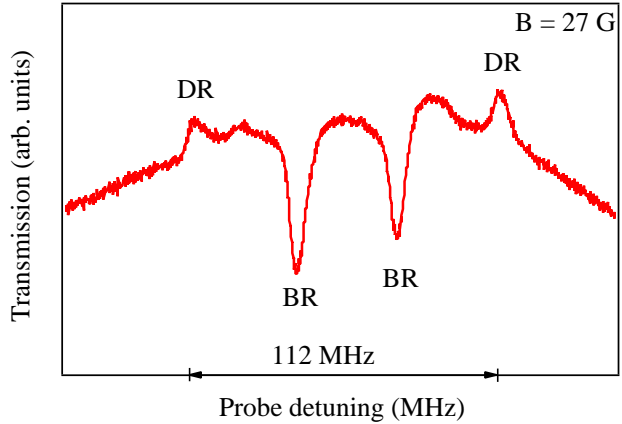} 
	\caption{Experimental spectrum of probe transmission showing both BRs and DRs. The curve is obtained with a probe power of \SI{200}{\micro\watt}, control power of 13.6 mW, and a magnetic field of 27 G.}
	\label{result_secondlambda}
\end{figure}

The experimental spectrum shown in Fig.~\ref{result_secondlambda}, is obtained with a probe power of \SI{200}{\micro\watt}. This is higher than that for the other system, and is chosen to get an adequate signal-to-noise ratio (SNR). The curve also shows the resonances are inverted, i.e.~DRs become BRs and BRs become DRs.

\subsubsection{Theoretical results}

\begin{figure}
	\centering
	\includegraphics[width=.6\textwidth]{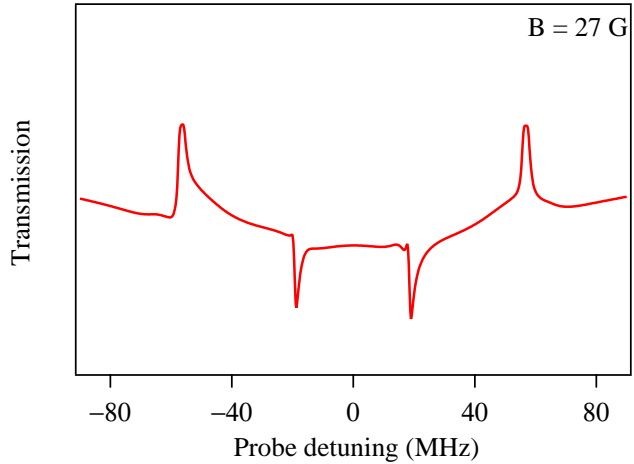}
	\caption{Calculated spectrum for the second $\Lambda$-type system. The Rabi frequency for the probe beam is $1.69 \, \Gamma_e$ (corresponding to the experimental value of \SI{200}{\micro \watt}), and that of the control beam is $13.9 \, \Gamma_e$ (corresponding to the experimental values of 13.6 mW). The values of the relaxation rates used are $\gamma_{\rm b} = 0.035 \,\Gamma_e$, $\gamma_{\rm d} = 0.000056 \,\Gamma_e$ , $\gamma_{\rm w} = 0.000174 \,\Gamma_e$.}
	\label{secondlambda_theory}
\end{figure}

The calculated spectrum is shown in Fig.~\ref{secondlambda_theory}. In this case, the probe beam corresponds to the $\vec{E}_1$ field, while the control beam corresponds to the $\vec{E}_2$ field. As in the other case, the calculation is done with probe and control Rabi frequencies corresponding to the ones used in the experiment. The values of the relaxation rates are the same as before, and are given in the figure caption for completeness. Interestingly, it reproduces the experimental observation that the resonances are inverted. As in the other case, the linewidth is narrower and the lineshape is asymmetric, probably for the same reasons as mentioned before.

\section{Conclusions}
In summary, we have studied EIT resonances in a Rb vapor cell with anti-relaxation coating on the walls and in the presence of a 27 G magnetic field. Both bright and dark resonances are obtained as the probe frequency is scanned. The sign of the resonances can be reversed by interchanging the transitions coupled by the probe and control beams. A theoretical explanation based on a density-matrix analysis of the sublevels involved in the transition reproduces the experimental spectra quite well. The different sign of the resonances implies that the same system can be used for both slow and fast propagation of light.

\section*{Acknowledgments}
A. S. and D. S. are grateful to S. Cartaleva for the ARC glass cell and to E. Mariotti for the PDMS material. V. B. acknowledges financial support from a D. S. Kothari  post-doctoral  fellowship  of  the  University  Grants Commission, India.

%\bibliographystyle{unsrt}%elsarticle-num
%\bibliography{eitrefs}

\end{document}